**Theory of Resonant and Nonresoant Magnetoelectric EffectsforLayered Composites with Anisotropic Piezoelectric Properties**

Deepak Rajaram Patil[1,a)], Yisheng Chai[1,a)], Rahul C. Kambale[2], Byung-Gu Jeon[1], Kyongjun Yoo[1], Jungho Ryu[2], Woon-Ha Yoon[2], Dong-Soo Park[2], Dae-Yong Jeong[3], Sang-Goo Lee[4], Jeongho Lee[4], Joong-Hee Nam[5], Jeong-Ho Cho[5], Byung-Ik Kim[5], and Kee Hoon Kim[1,b)]

[1]CeNSCMR, Department of Physics and Astronomy, Seoul National University, Seoul 151-747, Republic of Korea.
[2]Functional Ceramics Group, Korea Institute of Materials Science (KIMS), 66 Sangnam-Dong, Changwon, Gyeongnam 641–831, Republic of Korea.
[3]School of Materials Engineering, Inha University, Incheon 402-751, Korea.
[4]iBULe Photonics Co. Ltd, 7-39 Songdo-dong Yeonsu-gu Incheon, Korea.
[5]Center for Electronic Component Research, Korea Institute of Ceramic Engineering and Technology, Seoul 153-801, Republic of Korea

[a)]D. R. Patil and Y. S. Chai contributed equally to this work.
[b)] Electronic mail:khkim@phya.snu.ac.kr

**ABSTRACT:**

A new theory is developed for the magnetoelectric (ME) coupling in a symmetric 2-2 ME laminate having a piezoelectric crystal, particularly with anisotropic planar piezoelectric properties. Based on the average field method, the expressions for the transverse ME voltagecoefficientsare derivedat low and resonance frequencies. The resultant theoretical expressions predict that transverse ME voltage coefficients become anisotropic under in-plane magnetic fields at both low and resonance freuquencies. Furthermore, numerical simulations based on the material parameters of a representative symmetric 2-2 trilayer, composed of Ni/[011]-oriented Pb(Mg$_{1/3}$Nb$_{2/3}$)O$_3$–PbTiO$_3$ crystal/Ni, show the existence of multiple resonance frequencies and the characteristic phase difference in the complex ME voltages at each resonant frequency. All these thereoretical predictions are demonstrated to be in good agreement with theexperimental ME data both at low and resonant frequencies. The theory developed here could be broadly applicable to thevarious types of layered ME composites with any piezoelectrics with anisotropic piezoelectric coefficients.

**I.   INTRODUCTION:**

Multiferroic materials offer a unique capability of cross-control of polarization ($P$) by a magnetic field ($H$) or magnetization ($M$) by an electric field ($E$), which is termed as the direct or converse magnetoelectric (ME) effects, respectively.[1,2] The material system has drawn great research interest in recent years due to their promising application potential for novel multifunctional devices[3-6] as well as the fundametal scientific questions related to the spin frustration and the spin-orbit coupling.[7,8] The ME effect is usually quantified in terms of the ME coefficient $\alpha=\delta P/\delta H$ or the ME voltage coefficient $\alpha_E=\delta E/\delta H$, which satisfies the relation of $\alpha = \varepsilon_o\varepsilon_r\alpha_E$ where $\varepsilon_r$ is the relative permittivity of the material. Although there have been lots of new single phase materials showing greatly improved values, their operation temperature mostly remain below room termperatures.[7-12]

For practical applications at room temperature, significant research efforts have been made to utilizethe strain-induced ME coupling in the ME composites composed of magnetostrictive and piezoelectric maerials.[9-12] The ME effect in those ME composites is well known to occur as a combination of two step processes, i.e., magnetic-field-induced mechanical strain (magnetostriction) and stress-induced electric field generation (converse piezoelectric effect). Extensive experimental investigations have indeed focused on improving the ME coupling either by choosing a material combination with superior striction properties or by increasing interface coupling based on the different geomerical connectivity schemes (e.g., 0-3, 2-2, 1-3, and 1-2 structures).[9-12] The former example includes the approaches to utilize the metglas and $Pb(Mg_{1/3}Nb_{2/3})O_3–PbTiO_3$ (PMN-PT), which show superior magnetostriction/piezoelectric properties.[13-17] Combined with the former approach, the latter approach adopting the 1-2 structure using a piezofiber layer of PMN-PT sandwiched between two metglas layers has indeed produced the record high ME voltages of 45 V/cmOe at off-resonance and 1100 V/cmOe at resonance.[13]

Theoretically, several methods have been applied for predicting the ME voltages at both low and resonance frequencies, including average field, equivalent circuit and Green function methods.[18-28] In particular, Bichurin, Petrov, Srinivasan, and coworkers have pioneered to develop theoretical modeling of the ME effect at a low frequency region and carry out experimental ME effect measurements in bilayer and multilayer composites of ferrite–(Pb,Zr)TiO$_3$.[18,21,22] They derived the expressions for longitudinal ($\alpha_{E33}$), transverse ($\alpha_{E31}$) and in-plane longitudinal ($\alpha_{E11}$) ME coefficients, and predicted that ME voltage coefficients are dependent on the product of piezomagnetic and piezoelectric coefficients, interface coupling and volume fraction of the two phases. Moreover, the transverse ME coefficient was predicted to be generally bigger than the longitudinal one. All these theoretical considerations at low frequencies indeed matched fairly well with the experimental results.[18,21,22] It is also noteworthy that all the theories at low frequency regions are based on the constitutive equations for the strain components $S_i^m$ and $S_i^p$ ($i$=1 and 2) of the magnetic and piezoelectric phases, and that they take into account the Poisson's effect by considering the longitudinal and transverse dynamic strains simultaneously.[18,19]

Theoretical models for predicting the ME effects near the mechanical resonance have been also developed to successfully explain the greatly enhanced ME peak at the mechanical resonance.[19,24-26] However, the existing theories have so far considered the expressions of the resonant ME voltages for 1D isotropic laminate with length $l$ much larger than width $w$ so that they included only one dimentional (1D) internal stresses along the length direction while ignoring the transverse stresses along the width direction. As a result, the predicted ME voltage coefficient were generally biggerthan the experimental data. To remedy this effect, Bao and Luo[29] recently developed a new theoretical model for the resonant ME effects considering the 2D stresses. The theory naturally predicted two resonant frequencies coming from the stress components along the width and length, which could not be predicted by the former theories considering the 1 D stress. Moreover, the theoretical modeling based on 2D stresses are found to be in better agreement with the experimental data than that based on the 1D stress only.

We note that almost all the theoretical models for the ME effects at resonant and low frequencieshave indeed considered the isotropic magnetostrictive and piezoelectric phases, particularlywiththe isotropic in-plane striction properties. In particular, piezoelectric materials with isotropic piezoelectric coefficients ($d_{31}=d_{32}$) and elastic compliances ($s_{11}^p = s_{22}^p$)areadopted. As a result, the ME laminate compositegives rise to the isotropic stress and strain components alongthe in-plane directions andthus results in the isotropic transverse ME voltage coefficients i.e., $\alpha_{E31}= \alpha_{E32}$. In contrast, if the piezoelectric phase with anisotropic piezoelectric coefficients i.e., ($d_{31}\neq d_{32}$) and elastic compliances ($s_{11}^p \neq s_{22}^p$) are adopted, thetransverse stress and strain componentsbecome different along the orthogonal in-plane directions, resulting in the anisotropic transverse ME voltage coefficientsat both low and resonance frequencies. Recently, we have indeed reported the observation of the giant enhancement and strong anisotriopy in the transverse ME voltage coefficients inthesymmetric metglas/PMN-PT/metglaslaminate, upon using the [011]-oriented PMN-PT crystalwith anisotropic transverse piezoelectric properties.[30] However, a proper theoretical modeling is still lacking to properly understand the experimental findings such as anisotropic ME coupling effect depending on the in-plane magnetic field directions and resonance frequencies.

In this paper, we present the new theory for the ME coupling in a symmetric trilayer laminate with a piezoelectric crystal with anisotropic planar piezoelectric properties. Theoretical expressions for the transverse ME voltage coefficients are derived both at low and resonance frequencies by properly taking in to account the planar 2D stress effects. The simulation results based on the theoretical expression predicted two resonance frquencies and characteristic phase difference in the complex transverse ME voltage coefficients at each resonant frequency, being consistent with the experimental results. We also present several predictions of the theory such as anisotropic ME voltage signals in both low and resonant frequencies, and compare with the exeperimental ME data.

## II. THEORETICAL MODELING:

### 1. ME effect at low frequency:

We consider a symmetric layered composite structure composite of [011]-oriented PMN-PT crystal sandwitched between two Ni layers, which has the form of a thin plate with length $l$ and width $w$ as shown in Fig. 1(a). Here, we adopted a special case of [011]-oriented PMN-PT single crystal which exhibited anisotropic in-plane piezoelectric coefficients with a positive $d_{31}$ of 610 pC/N along $[0\bar{1}1]$ and a negative -1883 pC/N along [100] (Fig. 1(b)).[26] The anisotropy in piezoelectric coefficients could generate in-plane tensile stress along $[0\bar{1}1]$ and compressive stress along [100] while applying electric field parallel to [011] direction.[26]

The theory described here is based on the following equations for the strain components $S_i^m$ and $S_i^p$ of the magnetic and piezoelectric layers and the electric displacement $D_3$ of the piezoelectric layer, which can be written as,

$$S_1^p = s_{11}^p T_1^p + s_{21}^p T_2^p + d_{31}^p E_3$$

$$S_2^p = s_{12}^p T_1^p + s_{22}^p T_2^p + d_{32}^p E_3$$

$$S_1^m = s_{11}^m T_1^m + s_{21}^m T_2^m + q_{11}^m H_1$$

$$S_2^m = s_{12}^m T_1^m + s_{22}^m T_2^m + q_{12}^m H_1$$

$$D_3 = d_{31}^p T_1^p + d_{32}^p T_2^p + \varepsilon_{33}^T E_3 \tag{1}$$

Here, $T_i^p$ are the stress components in the piezoelectric phase, $s_{ij}^p$ are the compliance coefficients of the piezoelectric phase under constant stress, $T_i^m$ are the stress components in the magnetostrictive phase, $s_{ij}^m$ are the compliance coefficients of the piezoelectric phase under constant stress, $\varepsilon_{33}^T$ is the permittivity, $d_{31}^p$ and $d_{32}^p$ are the piezoelectric coefficients, $q_{11}^m$ and $q_{12}^m$ are the piezomagnetic coefficients, and $E_3$ and $H_1$ are the electric and magnetic field strengths.

Considering the internal stresses within magnetostrictive and piezoelectric plates with the perfect interface coupling i.e. $k = 1$ and according to Newton's third law, we get following relationships about the internal forces between layers:

$$fT_1^m + (1-f)T_1^p = 0, \quad fT_2^m + (1-f)T_2^p = 0 \tag{2}$$

Where, $f$ denotes the volume fraction of the magnetostrictive phase in the ME laminate and is equal to 0.5 in our laminates.

For the solutions of Eq. (2), the following boundary conditions were used:

$$T_1^p = -T_1^m, \quad T_2^p = -T_2^m, \quad S_1^p = S_1^m, \quad S_2^p = S_2^m \tag{3}$$

Combining Eqs. (1), (2) and (3), we can obtain

$$T_1^P = [\underline{s_{22}}(q_{11}^m H_1 - d_{31}^P E_3) - \underline{s_{21}}(q_{12}^m H_1 - d_{32}^P E_3)]/2[\underline{s_{11}}\underline{s_{22}} - \underline{s_{21}}\underline{s_{12}}]$$
$$T_2^P = [\underline{s_{11}}(q_{12}^m H_1 - d_{32}^P E_3) - \underline{s_{12}}(q_{11}^m H_1 - d_{31}^P E_3)]/2[\underline{s_{11}}\underline{s_{22}} - \underline{s_{21}}\underline{s_{12}}] \qquad (4)$$

where, the effective compliance coefficients were defined as,

$$\underline{s_{11}} = \frac{(s_{11}^p + s_{11}^m)}{2}, \quad \underline{s_{21}} = \underline{s_{12}} = \frac{(s_{21}^p + s_{21}^m)}{2}, \text{ and } \underline{s_{22}} = \frac{(s_{22}^p + s_{22}^m)}{2}$$

Magnetoelectric coupling is estimated from the induced field $\delta E$ across the sample that is subjected to an ac magnetic-field $\delta H$ in the presence of a bias field $H_{dc}$. Basic relations for transverse ME coefficients are obtained for two orientations of $H_{dc}$ and $\delta H$ i.e. along direction 1 or along direction 2. From Eqs. (1) and (4), and considering the open circuit condition $D_3=0$, the transverse ME coefficient $\alpha_{E31}= \delta E_3/\delta H_1$ can be expressed as,

$$\alpha_{E31} = \frac{(d_{31}^p \underline{s_{22}} q_{11}^m + d_{32}^p \underline{s_{11}} q_{12}^m) - \underline{s_{21}}(d_{31}^p q_{12}^m + d_{32}^p q_{11}^m)}{2\varepsilon_{33}^p[\underline{s_{11}}\underline{s_{22}} - (\underline{s_{21}})^2] - [\underline{s_{22}}(d_{31}^p)^2 + \underline{s_{11}}(d_{32}^p)^2 - 2\underline{s_{21}} d_{31}^p d_{32}^p]} \qquad (5)$$

Similarly, the transverse ME coefficient $\alpha_{E32}= \delta E_3/\delta H_2$ can be expressed as,

$$\alpha_{E32} = \frac{(d_{31}^p \underline{s_{22}} q_{21}^m + d_{32}^p \underline{s_{11}} q_{22}^m) - \underline{s_{21}}(d_{31}^p q_{22}^m + d_{32}^p q_{21}^m)}{2\varepsilon_{33}^p[\underline{s_{11}}\underline{s_{22}} - (\underline{s_{21}})^2] - [\underline{s_{22}}(d_{31}^p)^2 + \underline{s_{11}}(d_{32}^p)^2 - 2\underline{s_{21}} d_{31}^p d_{32}^p]} \qquad (6)$$

One can check the validity of the above equations by solving the above equations for isotropic 2D laminate with isotropic piezoelectric properties. After putting the isotropic piezoelectric properties, $s_{11}^p = s_{22}^p, s_{12}^p = s_{21}^p, d_{31}^p = d_{32}^p$ in the Eqs. (5) and (6) one can get,

$$\alpha_{E31} = -\frac{f(1-f)d_{31}(q_{11} + q_{21})}{\varepsilon_{33}\underline{s_{11}} - 2kfd_{31}^2} \qquad (7)$$

, where, $\underline{s_{11}} = f(s_{11}^p + s_{12}^p) + (1-f)(s_{11}^m + s_{12}^m)$. The Eq. (7) is in good agreement with the theory derived by Bichurin *et.al.*[18] for the isotropic ME laminate. Note that $\alpha_{E31}=\alpha_{E32}$ holds in the isotropic piezoelectric/magnetorictive media as the equality relations of $d_{31}=d_{32}$, $q_{11}=q_{22}$ and $q_{12}=q_{21}$ are valid.

## 2. ME effect at resonance frequency:

The ME effect in the composites is driven by the mechanical coupling between the piezoelectric and magnetic phases, the ME effect would be greatly enhanced when the piezoelectric or magnetic phase undergoes mehcanical resonance i.e., an electromechanical resonance(EMR) for the piezoelectric phase and ferromagnetic resonance (FMR) for the magnetic phase. Since the ME equations for $\alpha_{E31}$ and $\alpha_{E32}$ at low frequency show different expressions for different in-plane $H$, one can expect the same different form of expressions for $\alpha_{E31}$ and $\alpha_{E32}$ at resonance conditions. Therefore it is equally important to understand the theoretical modelling for the transverse ME voltage coefficient at resonance by solving the fundamental constitutive equations. In order to describe the ME volate coefficient at resonance conditions we need the equations of elastodynamics along with the above fundamental constitutive equations.

From Eqs.(1) and (3) the following equations for the effective parameters for the composites can be derived:

$$S_1^p = s_{11}^{p'}T_1^p + q_{11}^{m'}H_1 + d_{31}^{p'}E_3,$$

$$S_2^p = s_{22}^{p'}T_2^p + q_{12}^{m'}H_1 + d_{32}^{p'}E_3,$$

$$S_1^m = s_{11}^{m'}T_1^m + d_{31}^{p'}E_3 + q_{11}^{m'}H_1, \quad S_2^m = s_{22}^{m'}T_2^m + d_{32}^{p'}E_3 + q_{12}^{m'}H_1,$$

$$T_1^P = \frac{(s_{22}^p S_1^P - s_{21}^p S_2^P)}{(s_{11}^p s_{22}^p - s_{21}^p s_{12}^p)} - \frac{E_3(s_{22}^p d_{31}^P - s_{21}^p d_{32}^P)}{(s_{11}^p s_{22}^p - s_{21}^p s_{12}^p)}$$

$$T_2^P = \frac{(s_{12}^p S_1^P - s_{11}^p S_2^P)}{(s_{21}^p s_{12}^p - s_{11}^p s_{22}^p)} - \frac{E_3(s_{12}^p d_{31}^P - s_{11}^p d_{32}^P)}{(s_{21}^p s_{12}^p - s_{11}^p s_{22}^p)}$$

$$\begin{aligned}D_3 &= d_{31}^p T_1^p + d_{32}^p T_2^p + \varepsilon_{33}^T E_3 \\ &= [\frac{d_{31}^p(s_{22}^p S_1^P - s_{21}^p S_2^P)}{(s_{11}^p s_{22}^p - s_{21}^p s_{12}^p)} - \frac{d_{31}^p E_3(s_{22}^p d_{31}^P - s_{21}^p d_{32}^P)}{(s_{11}^p s_{22}^p - s_{21}^p s_{12}^p)}] + [\frac{d_{32}^p(s_{11}^p S_2^P - s_{12}^p S_1^P)}{(s_{11}^E s_{22}^E - s_{21}^E s_{12}^E)} - \frac{d_{32}^p E_3(s_{11}^p d_{32}^P - s_{12}^p d_{31}^P)}{(s_{11}^p s_{22}^p - s_{21}^p s_{12}^p)}] + \varepsilon_{33}^T E_3 \\ &= [\frac{S_1^P(d_{31}^p s_{22}^p - d_{32}^p s_{12}^p) - S_2^P(d_{31}^p s_{21}^p - d_{32}^p s_{11}^p)}{(s_{11}^p s_{22}^p - s_{21}^p s_{12}^p)}] + E_3(\varepsilon_{33}^T - \frac{s_{22}^p(d_{31}^P)^2 + s_{11}^p(d_{32}^P)^2 - (s_{12}^p + s_{21}^p)d_{31}^p d_{32}^p}{(s_{11}^p s_{22}^p - s_{21}^p s_{12}^p)})\end{aligned}$$

(8) where,

$$s_{11}^{p'} = \frac{s_{11}^p s_{21}^m - s_{11}^m s_{12}^p}{s_{21}^m + s_{12}^p}, \quad s_{22}^{p'} = \frac{s_{12}^m s_{22}^p - s_{12}^p s_{22}^m}{s_{12}^m + s_{12}^p}, \quad d_{31}^{p'} = d_{31}^p \frac{s_{21}^m}{s_{21}^m + s_{12}^p}, \quad d_{32}^{p'} = d_{31}^p \frac{s_{12}^m}{s_{12}^m + s_{12}^p},$$

$$s_{11}^{m'} = \frac{s_{11}^m s_{12}^p - s_{11}^p s_{21}^m}{s_{12}^p + s_{21}^m}, \quad s_{22}^{m'} = \frac{s_{12}^p s_{22}^m - s_{11}^m s_{22}^p}{s_{12}^p + s_{12}^m}, \quad q_{11}^{m'} = q_{11}^m \frac{s_{12}^p}{s_{12}^p + s_{12}^m}, \quad q_{12}^{m'} = q_{12}^m \frac{s_{12}^p}{s_{12}^p + s_{12}^m}.$$

Based on the coordinate system shown in **Fig.**1(a), and applying Newton's second law to the ME element, the equations of motion for any mass element oriented in the direction of the *x* and *y* axis can be respectively written as,

$$\frac{\partial^2 u}{\partial^2 x} + k_x^2 u = 0,$$

$$\frac{\partial^2 v}{\partial^2 y} + k_y^2 v = 0 \tag{9}$$

Where $u$, $v$ are the displacements of the mass element for the magnetoelectric element along $x$ and $y$ axis. The wave numbers $k_x$ and $k_y$ is given by:

$$k_x = \omega / \sqrt{(\frac{1-f}{s_{11}^p} + \frac{f}{s_{11}^m})/\overline{\rho})} \quad k_y = \omega / \sqrt{(\frac{1-f}{s_{22}^p} + \frac{f}{s_{22}^m})/\overline{\rho})}$$

$\overline{\rho} = f\rho_m + (1-f)\rho_p$ is the average mass density, where $\rho_p$ and $\rho_m$ represent the densities of the piezoelectric and magnetostrictive materials, respectively.

The solutions of Eq. (9) can be written as:

$$u = A_u \sin(k_x u) + B_u \cos(k_x u)$$
$$v = A_v \sin(k_y u) + B_v \cos(k_y u) \quad (10)$$

To determine $A_u$, $A_v$, $B_u$ and $B_v$, according to Eq. (8):

$$S_1^p = \frac{\partial u}{\partial x} = s_{11}^{p'} T_1^p + q_{11}^{m'} H_1 + d_{31}^{p'} E_3, \quad (11)$$

$$S_2^p = \frac{\partial v}{\partial y} = s_{22}^{p'} T_2^p + q_{12}^{m'} H_1 + d_{32}^{p'} E_3 \quad (12)$$

In free boundary conditions, when $u = 0$ or $l$ and $v = 0$ or $w$, $T_1^p$ and $T_2^p$ will become zero, which gives:

$$B_u = \frac{q_{11}^{m'} H_1 + d_{31}^{p'} E_3}{k_x}, B_v = \frac{q_{12}^{m'} H_1 + d_{32}^{p'} E_3}{k_x}$$

$$A_u = B_u \frac{\cos(k_x l) - 1}{\sin(k_x l)}, A_v = B_v \frac{\cos(k_y w) - 1}{\sin(k_y w)} \quad (13)$$

Putting Eq. (13) back to Eq. (10) and taking derivative, one can obtain:

$$S_1^p = \frac{\partial u}{\partial x} = (q_{11}^{m'} H_1 + d_{31}^{p'} E_3) \frac{[\sin(k_x(l-x)) + \sin(k_x x)]}{\sin(k_x l)}$$

$$S_2^p = \frac{\partial v}{\partial y} = (q_{12}^{m'} H_1 + d_{32}^{p'} E_3) \frac{[\sin(k_y(w-y)) + \sin(k_y y)]}{\sin(k_y w)} \quad (14)$$

The ME voltage coefficient is determined using the open circuit condition:

$$\int_0^w \int_0^l D_3 dx dy = 0 \quad (15)$$

Substituting Eq. (8) into (15):

$$\int_0^w\int_0^l D_3 dxdy = \int_0^w\int_0^l \left[\frac{S_1^P(d_{31}^P s_{22}^P - d_{32}^P s_{12}^P) - S_2^P(d_{31}^P s_{21}^P - d_{32}^P s_{11}^P)}{(s_{11}^P s_{22}^P - s_{21}^P s_{12}^P)}\right] + E_3\left(\varepsilon_{33}^T - \frac{s_{22}^P(d_{31}^P)^2 + s_{11}^P(d_{32}^P)^2 - (s_{12}^P + s_{21}^P)d_{31}^P d_{32}^P}{(s_{11}^P s_{22}^P - s_{21}^P s_{12}^P)}\right) dxdy$$

$$= \left(\left[\frac{(d_{31}^P s_{22}^P - d_{32}^P s_{12}^P)\int_0^w\int_0^l S_1^P dxdy - (d_{31}^P s_{21}^P - d_{32}^P s_{11}^P)\int_0^w\int_0^l S_2^P dxdy}{(s_{11}^P s_{22}^P - s_{21}^P s_{12}^P)}\right] + lwE_3\left(\varepsilon_{33}^T - \frac{s_{22}^P(d_{31}^P)^2 + s_{11}^P(d_{32}^P)^2 - (s_{12}^P + s_{21}^P)d_{31}^P d_{32}^P}{(s_{11}^P s_{22}^P - s_{21}^P s_{12}^P)}\right)\right) \quad (16)$$

Substituting $S_1^P$ and $S_2^P$ terms from Eq. (14) into (16):

$$\int_0^w\int_0^l S_1^P dxdy = (q_{11}^{m'}H_1 + d_{31}^{p'}E_3)w\int_0^l \frac{[\sin(k_x(l-x)) + \sin(k_x x)]}{\sin(k_x l)}dx = \frac{2(q_{11}^{m'}H_1 + d_{31}^{p'}E_3)w}{k_x \cot\frac{k_x l}{2}}$$

$$\int_0^w\int_0^l S_2^P dxdy = (q_{12}^{m'}H_1 + d_{32}^{p'}E_3)l\int_0^w \frac{[\sin(k_y(w-y)) + \sin(k_y y)]}{\sin(k_y w)}dy = \frac{2(q_{12}^{m'}H_1 + d_{32}^{p'}E_3)l}{k_y \cot\frac{k_y w}{2}} \quad (17)$$

Finally, the transverse ME coefficient $\alpha_{E31} = \delta E_3/\delta H_1$ is given by:

$$\alpha_{E31} = \frac{\left(\frac{2q_{11}^{m'}(d_{32}^P s_{12}^P - d_{31}^P s_{22}^P)}{k_x l \cot(k_x l/2)} + \frac{2q_{12}^{m'}(d_{31}^P s_{21}^P - d_{32}^P s_{11}^P)}{k_y w \cot(k_y w/2)}\right)}{\left(\frac{2d_{31}^{p'}(d_{31}^P s_{22}^P - d_{32}^P s_{12}^P)}{k_x l \cot(k_x l/2)} + \frac{2d_{32}^{p'}(d_{32}^P s_{11}^P - d_{31}^P s_{21}^P)}{k_y w \cot(k_y w/2)}\right) + [(s_{11}^P s_{22}^P - s_{12}^P s_{21}^P)\varepsilon_{33}^T - s_{22}^P(d_{31}^P)^2 - s_{11}^P(d_{32}^P)^2 + (s_{12}^P + s_{21}^P)d_{31}^P d_{32}^P]} \quad (18)$$

Similarly, the transverse ME coefficient $\alpha_{E32} = \delta E_3/\delta H_2$ is given by

$$\alpha_{E32} = \frac{\left(\frac{2q_{21}^{m'}(d_{32}^P s_{12}^P - d_{31}^P s_{22}^P)}{k_x l \cot(k_x l/2)} + \frac{2q_{22}^{m'}(d_{31}^P s_{21}^P - d_{32}^P s_{11}^P)}{k_y w \cot(k_y w/2)}\right)}{\left(\frac{2d_{31}^{p'}(d_{31}^P s_{22}^P - d_{32}^P s_{12}^P)}{k_x l \cot(k_x l/2)} + \frac{2d_{32}^{p'}(d_{32}^P s_{11}^P - d_{31}^P s_{21}^P)}{k_y w \cot(k_y w/2)}\right) + [(s_{11}^P s_{22}^P - s_{12}^P s_{21}^P)\varepsilon_{33}^T - s_{22}^P(d_{31}^P)^2 - s_{11}^P(d_{32}^P)^2 + (s_{12}^P + s_{21}^P)d_{31}^P d_{32}^P]} \quad (19)$$

The above equations correspond to a special case of the ME theory in which the anisotropic piezoelectric properties ($s_{11}^P \neq s_{22}^P, s_{12}^P = s_{21}^P$ and $d_{31}^P \neq d_{32}^P$) are considered with $l \neq w$. One can check the validity of the above equations by solving the above equations for isotropic 2D laminate with isotropic piezoelectric properties. After putting the isotropic piezoelectric material parameters $s_{11}^P = s_{22}^P, s_{12}^P = s_{21}^P, d_{31}^P = d_{32}^P$ in the Eqs. (18) and (19) one can get,

$$\alpha_{E31} = -\frac{\left(\dfrac{q_{11}^{m'}}{k_x l \cot(k_x l/2)} + \dfrac{q_{12}^{m'}}{k_y w \cot(k_y w/2)}\right)}{\left(\dfrac{d_{31}^{p'}}{k_x l \cot(k_x l/2)} + \dfrac{d_{31}^{p'}}{k_y w \cot(k_y w/2)}\right) + \dfrac{\varepsilon_{33}^T}{X}[1 - \dfrac{X(d_{31}^p)}{\varepsilon_{33}^T}]} \quad (20)$$

with $X = \dfrac{2 d_{31}^p}{(s_{11}^P + s_{12}^P)}$

The above equation is in good agreement with the theory derived by Bao and Luo[27] for the 2D isotropic ME laminate. Note that the coordinate system and notation of the physical terms in above equations have been changed from the original version[27] to be consistent with our case.

Similarly one can find the equation for 1D isotropic laminate with length $l$ much larger than width $w$ and thickness $t$,[9]

$$\alpha_{E31} = \left(\frac{d_{31}^p q_{11}^m s_{12}^p \tan(kl/2)}{s_2[2(d_{31}^p)^2 - (s_{11}^p + s_{12}^p)\varepsilon_{33}^T]kl - (d_{31}^p)^2 s_{12}^m \tan(kl/2)}\right) \quad (21)$$

with $s_2 = \dfrac{s_{12}^m + s_{12}^p}{2}$. The above Eq. (21) is not fully consistent with the reported one in ref. [9] if we assume $f = 0.5$ and the effective permeability $\mu = 1$. This implies that the transverse stress should be considered even though the sample is 1D like.

From Eqs. (18) and (19), we can clearly find two different resonance frequencies likely due to the difference in the elastic compliances ($s_{11}^P \neq s_{22}^P$) of the piezoelectric materials.

$$f_{r1} = \frac{1}{2l}\sqrt{(\frac{1-f}{s_{11}^p} + \frac{f}{s_{11}^m})/\overline{\rho})}$$

$$f_{r2} = \frac{1}{2l}\sqrt{(\frac{1-f}{s_{22}^p} + \frac{f}{s_{11}^m})/\overline{\rho})} \quad (22)$$

## III. NUMERICAL SIMULATION OF THEORY

As an example, numerical estimations of ME voltage coefficient are considered for a trilayer of Ni/PMN-PT/Ni based on the above theories. We have performed the numerical calculations for the transverse ME voltage coefficient both at low and resonance frequency. For low frequency, the calculations were done by putting the the material parameters given in Table 1 into Eqs. (5) and (6). The calculated values of $\alpha_{E31}$ and $\alpha_{E32}$ are found to be -3.451 and 4.572 V/cmOe, respectively, showing opposite signs and differnt magnitudes with each other. Morover, the anisotropy ratio is found to be ~0.75. The anisotropy in the $\alpha_{E31}$ and $\alpha_{E32}$ can not be simply explained by the previously reported theories for isotropic laminates. Here, the vaules of material parameter used are not ideal therefore the relative ratio is very sensitive to the values of materials parameters. We have indeed found that the slight difference in the vaules give drastic change in anisotropy ratio.

The frequency dependence of $\alpha_{E31}$ and $\alpha_{E32}$ were calculated by using Eqs. (18) and (19) by considering the square symmetric laminate with $l = w$. The numerical simulations of Eqs. (18) and (19) were given in the Figs. 2 and 3. It should be noted that, the energy loss has not been considered in the present theoretical model, which are present in the real ME structure. These losses determine the resonance line width and the maximum value of the magnetoelectric coefficient. To take into account the energy loss, we set $\omega$ equal to $\omega' - i\omega''$ with $\omega''/\omega' = 10^{-3}$. Figure 2 shows the simulation results of modulus of ME voltage coefficient, i.e. $|\tilde{\alpha}_E|$ as a function of frequency for $H//[0\bar{1}1]$ and $H//[100]$. Two sharp peaks were observed at $f_1 = 176$ kHz and $f_2 = 205$ kHz for both the $H$ directions. Moreover, the relative magnitude of each $|\tilde{\alpha}_E|$ peak is different for two $H$ directions, consisternt with the low frequency data. On the other hand, the ratio of $|\alpha_{E31}/\alpha_{E32}|$ was overall frequency dependent as shown in the inset of Fig. 2, showing drastic changes near the resonant frequencies. For instance, the ratio at frequency $f_1$ is ~0.62 while it is ~1.5 at $f_2$. The present theoretical model not only predicts the anisotropy in the magnitude of $\tilde{\alpha}_E$ but also provides useful information about the relative phases between transverse ME voltage coefficients. The complex behavior of $\tilde{\alpha}_E$ is shown in the Fig. 3. The real and imaginary components of $\tilde{\alpha}_E$ generally exhibits the characteristic features as shown in Fig. 3 (a) and 3(b) for $H//[0\bar{1}1]$ and $H//[100]$, respectively. Most importantly, the phase change of 180 degrees was observed for $H$ changing from $[0\bar{1}1]$ to [100] direction. The anisotropy in both the sign and magnitde of transverse ME voltage coefficients is very unique and never been considered before. However, the experimental verification is need to check the validity of the present theoretical expressions.

## IV. EXPERIMENTAL RESULTS AND COMPARISON WITH NUMERICAL SIMULATIONS

### 1. EXPERIMENTALS

To verify the proposed theoretical models, we have compared the numerical simulations with the experimental data. Measurements were done on the symmetric Ni/PMN-PT/Ni laminates with lateral dimensions of 10×10 mm$^2$ by use of [011] oriented single crystals of PMN-PT. High quality PMN-PT (0.7Pb(Mg$_{1/3}$Nb$_{2/3}$)O$_3$–0.3PbTiO$_3$) single crystals were grown by the Bridgeman method (IBULE Photonics, Korea) and cut in a planar shape with a thickness of 0.3 mm oriented along [011] direction (Fig. 1(b)). The ME laminates were prepared by stacking six Ni layers (Alfa Aesar, 99.9%), of which thickness is 0.15 mm on the top and bottom surfaces of the PMN-PT using a silver epoxy. The samples were poled by applying a dc electric field of 10 kV/cm. To investigate quantitatively the ME coupling, a magnetoelectric susceptometer, working at both resonant and low frequency conditions, was used. A pair of Helmholtz coils was used to generate AC magnetic field $\delta H_{ac}$ in a broad frequency range ($f$ =194 Hz–1 MHz) and the resultant AC voltage across the sample was measured by a lock-in amplifier as a function of $H_{dc}$ to estimate a complex ME voltage coefficient $\tilde{\alpha}_E \equiv \alpha_E + i\,\text{Im}(\tilde{\alpha}_E)$. The material parameters for each phases have been given in the following Table 1.

## 2. EXPERIMENTAL RESULTS AND COMPARISON

Figure 4 presents $\alpha_E$ curves of the laminate with [011]-cut crystal measured at $f$=194 Hz for $H_{dc}$//[100] and //[$0\bar{1}1$]. For both $H_{dc}$ directions, $\alpha_E$ exhibits a typical $H_{dc}$ dependence showing a sign change with respect to the reversal of $H_{dc}$ direction. Moreover, $\alpha_E$ values along H//[$0\bar{1}1$] and H//[100] clearly show opposite signs and the maximum magnitudes become 0.64 and 1.92 V/cmOe, respectively, constituting their ratio of ~0.34, being consistent with the theoretical calculations. The theory and the experiments are in good agreement with each other. The difference between expected value and the experimental data might be due to the difference between parameters used in theory and those in our real samples.

The frequency dependence of $|\tilde{\alpha}_E|$ for the laminate with [011]-oriented PMN-PT (Fig. 5), finding two strong $|\tilde{\alpha}_E|$ peaks at $f_1$ = 183 kHz and $f_2$ = 217 kHz for both $H$//[$0\bar{1}1$] and $H$//[100]. The magnitude of each $|\tilde{\alpha}_E|$ peak is different for the two $H$ directions, showing consistecy with the numerical simulation data (Fig.2). Moreover, similar to the simulation data, the phase difference of 180 degree is ovserved between $\tilde{\alpha}_E$ for different $H$ directions, as shown in Fig. 6(a) and 6(b), respectively. The experimental results show good agreement with the theoretical simulations. However, the experimental ratio |$\alpha_{E31}$/ $\alpha_{E32}$| is alomost same over the

wide frequency range (Inset of Fig. 5), showing inconsistency with the theoretical calculations. The difference in the theoretical ratio over wide frequency range can be expected as it is strongly influenced by the small changes in the material parameters. The material parameters used in the present calculations are not ideal while they might be different for our used PMN-PT and Ni.

## V.  CONCLUSION

In conclusion, a new theory for the transverse ME voltage coefficients at low and resonance frequency were derived. Our theory employed anisotropic transverse piezoelectric properteis of the piezoelectric phase predicting different equations of transverse ME voltage coefficients for different in-plane magnetic fields. The numerical simulations show multiple resonance frequencies and phase differenece between transverse ME voltage coefficients. The theoretical results show good agreement with the experimental results. The present theory is likely to provide unique tool to pursuing investigations on ME lamiates with different anisotropic piezoelectric phases. This theory provides theoretical understanding to optimise different material parameters in order to achieve enhancement in the ME voltage coefficients.

**FIGURE CAPTIONS:**

**FIG. 1.** (a) A thin ME laminate and defined Cartesian coordinates and (b) Schematic of PMN-PT single crystals with [011]-orientation. Out of 8 possible polarization directions inherent to the rhombohedral symmetry, 2 polarization directions are chosen for Cut-Bupon poling (red dotted lines).

**FIG. 2.** Numerical simulation of frequency dependence of the modulus of ME voltage coefficient i.e., $|\alpha_E|$ along different $H$ directions for the Ni/[011]-PMN-PT/Ni laminate.

**FIG. 3.** Numerical simulation of frequency dependence of $\alpha_E$ and $\mathrm{Im}(\tilde{\alpha}_E)$ along (a) $H//[0\bar{1}1]$ and (b) $H//[100]$ for the Ni/[011]-PMN-PT/Ni laminate.

**FIG. 4.** $H_{dc}$ dependence of $\alpha_E$ at a frequency $f=194$ Hz for Ni/[011]-PMN-PT/Ni laminates.

**FIG. 5.** Experimental results of frequency dependence of the modulus of ME voltage coefficient i.e., $|\alpha_E|$ along different $H$ directions for the Ni/[011]-PMN-PT/Ni laminate.

**FIG. 6.** Experimental results of frequency dependence of $\alpha_E$ and $\mathrm{Im}(\tilde{\alpha}_E)$ along (a) $H//[0\bar{1}1]$ and (b) $H//[100]$ for the Ni/[011]-PMN-PT/Ni laminate.

**Table 1:** Material parameters for Ni and [011]-oriented PMN-PT single crystal used for theoretical modelling.

| Materials | $s_{11}^p$ or $s_{11}^m$ ($10^{-12}$ m²/N) | $s_{22}^p$ or $s_{22}^m$ ($10^{-12}$ m²/N) | $s_{12}^p$ or $s_{12}^m$ ($10^{-12}$ m²/N) | $q_{11}$ ($10^{-9}$ m/A) or $d_{31}$ ($10^{-12}$ C/N) | $q_{22}$ ($10^{-9}$ m/A) or $d_{32}$ ($10^{-12}$ C/N) | $q_{12}$ ($10^{-9}$ m/A) | $\varepsilon_{33}/\varepsilon_0$ |
|---|---|---|---|---|---|---|---|
| Ni[a] | 4.57 | - | -1.37 | 1.25 | - | -0.59 | - |
| [011]PMNPT[b] | 18 | 112 | -31.1 | 610 | -1883 | - | 4003 |

[a]Cited from Ref. 31, 32.
[b]Cited from Ref. 23.

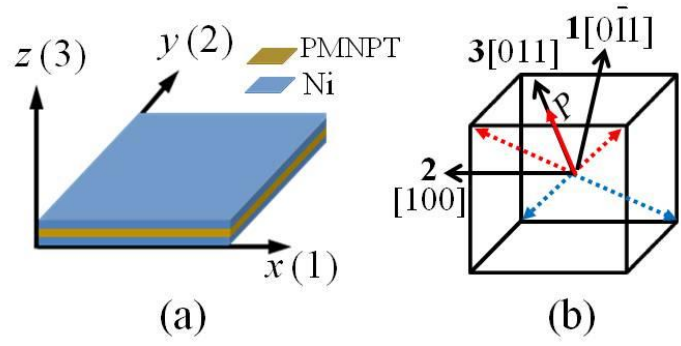

**FIG. 1.**

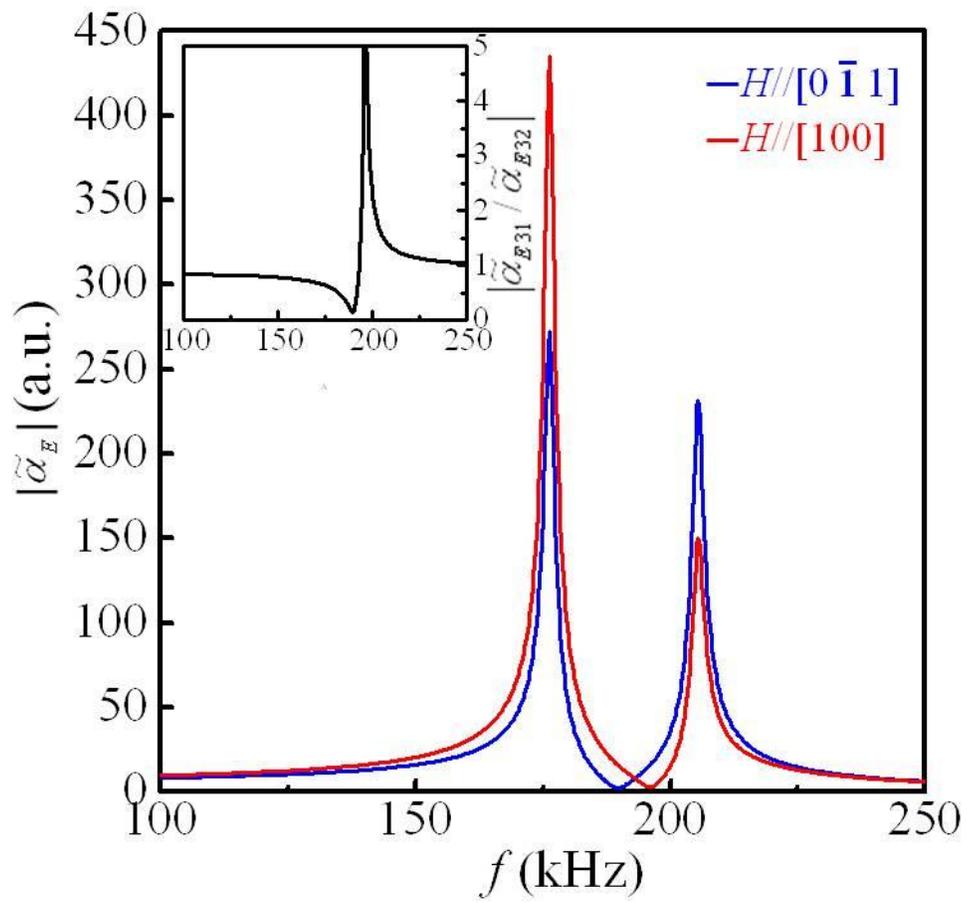

**FIG. 2.**

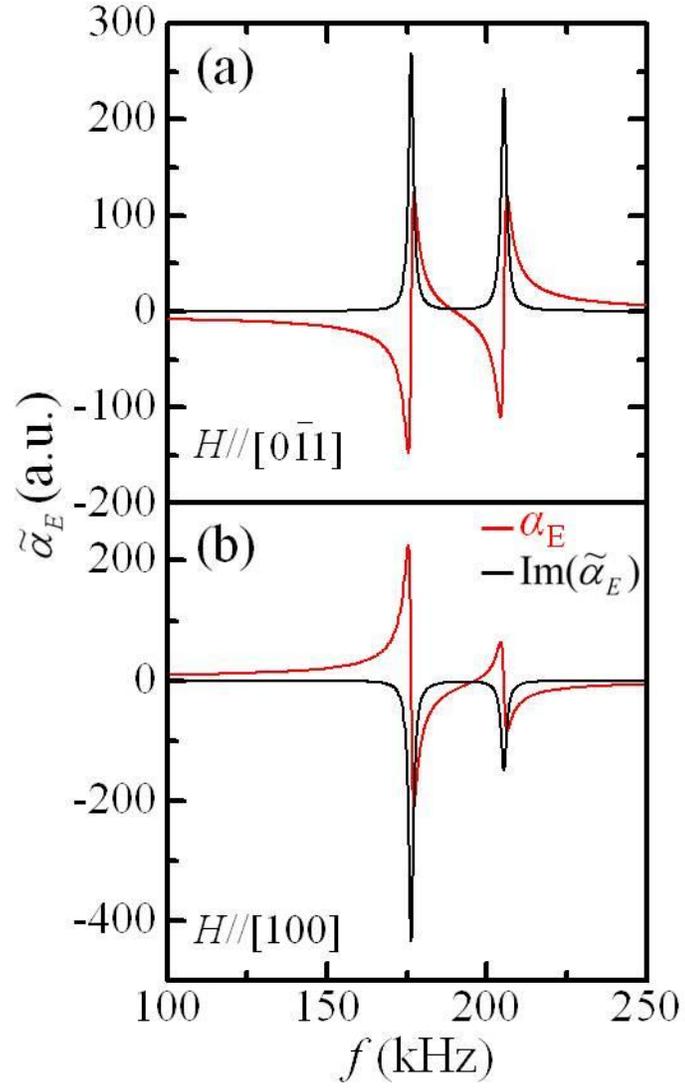

**FIG. 3**

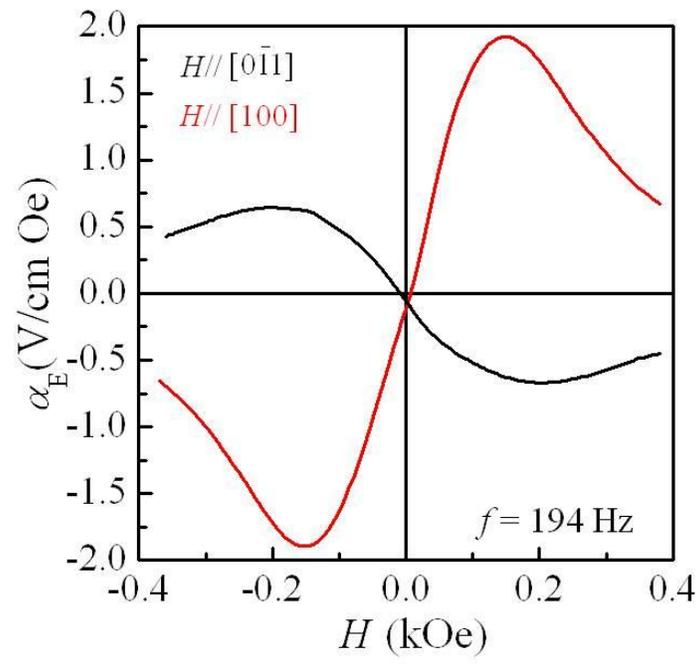

**FIG. 4.**

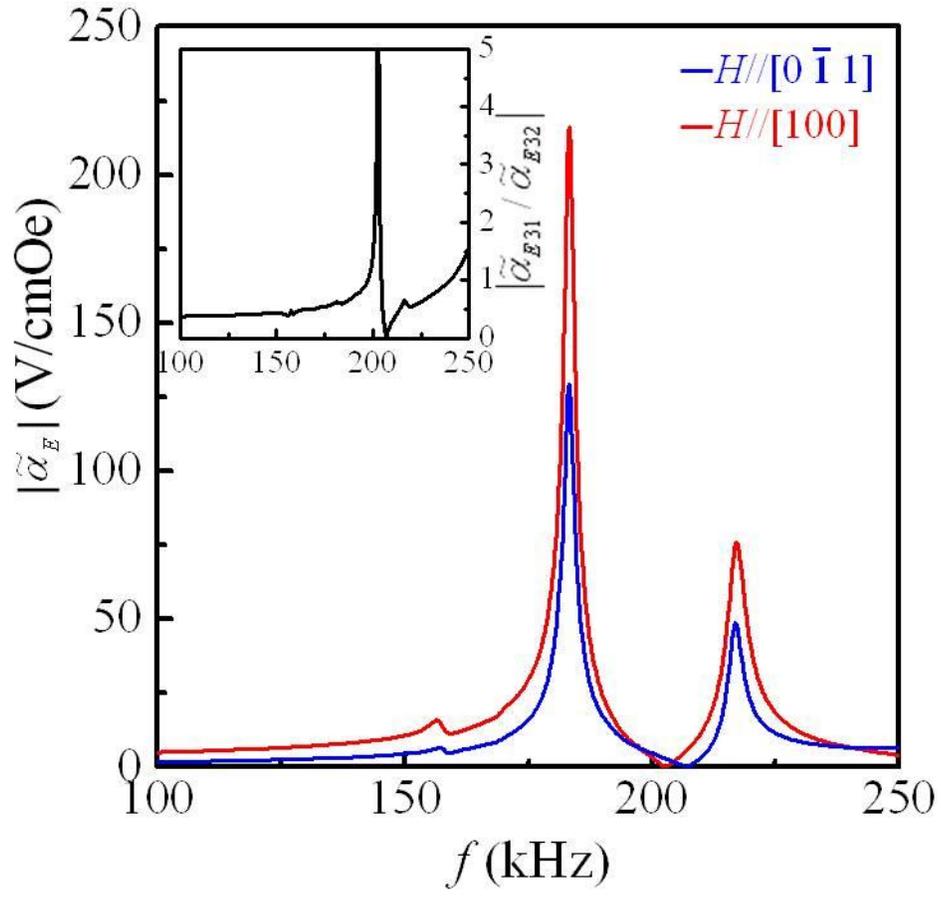

**FIG. 5.**

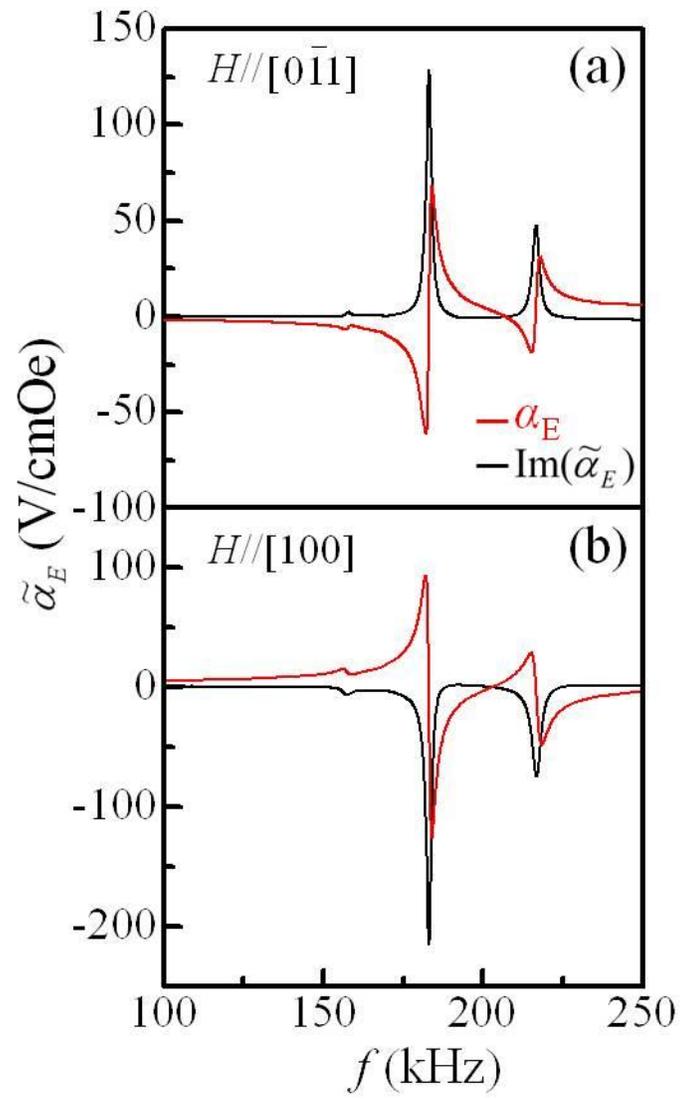

**FIG. 6.**